\documentclass[twocolumn,appendixfloats,tighten]{aastex6}
\usepackage{newtxtext,newtxmath}
\usepackage[T1]{fontenc}
\usepackage{ae,aecompl}

\usepackage{amsmath}	
\usepackage{amssymb}	

\usepackage{ctable}
\usepackage{url}
\usepackage{xspace}
\usepackage[normalem]{ulem}
\usepackage{hyperref}
\usepackage[all]{hypcap}
\usepackage{graphicx}

\def\msun{{\rm\,M_\odot}}

\def\msun{{\rm\,M_\odot}}

\newcommand{\be}{\begin{equation}}
\newcommand{\ee}{\end{equation}}

\def\h2{${\rm\,H_2}$}

\usepackage{color}

\begin{document}

\title{Being careful with the field formation interpretation of GW190412}

\author{Mohammadtaher Safarzadeh\altaffilmark{1}, Kenta Hotokezaka\altaffilmark{2,3}}
\altaffiltext{1}{Center for Astrophysics | Harvard \& Smithsonian, 60 Garden Street, Cambridge, MA, USA
\href{mailto:msafarzadeh@cfa.harvard.edu}{msafarzadeh@cfa.harvard.edu}}
\altaffiltext{2}{Research Center for the Early Universe, Graduate School of Science, University of Tokyo, Bunkyo-ku, Tokyo 113-0033, Japan}
\altaffiltext{3}{Department of Astrophysical Sciences, Princeton University, Princeton, 4 Ivy Ln, NJ 08544, USA}

\begin{abstract}
The LIGO/Virgo Scientific Collaboration recently announced the detection of a compact object binary merger, GW190412, as the first asymmetric binary black hole (BBH) merger with mass ratio  $q\approx0.25$.
Other than the mass ratio, this BBH has shown to have a positive effective spin of around $\chi_{\rm eff}\approx0.28$. 
Assuming a field formation channel, associating this effective spin to either the primary or the secondary BH each has its implications:
If the spin of the BBH comes form the primary BH, it has consequences for the efficiency of angular momentum transport in the formation of the BH.
If, on the other hand, the spin is due to the secondary BH through tidal spin-up processes, 
one has to note that (i) such processes have very short delay-times, and (ii) subsequently, their local merger rate is determined by local star formation rate at assumed formation metallicity of the BBH. 
We show that the predicted merger rate density from this channel is $\lesssim 0.3~\rm Gpc^{-3} yr^{-1}$ and in tension with the rather high local merger rate of such systems which we  estimate from this single event to be $\sim 1.7^{+2.5}_{-1.4}~\rm Gpc^{-3} yr^{-1}$ (90\% confidence interval, and assuming 50 days of observing time). Large natal kicks ($v\gtrsim 500\,{\rm km/s}$) would be required to get such BBHs with an in-plane spin component to account for the marginal detection of precession in GW190412. However, this would only exacerbate the tension as the estimated local merger rate would be further decreased. Similarly, the formation of such systems through the dynamical assembly is exceedingly rare, leaving this system a dilemma hard to account for with the currently accepted paradigms of BBH formation. 
\end{abstract}

\section{Introduction}
\label{sec:intro}

The spin distribution of the black holes at birth is largely unknown. 
The majority of the LIGO-Virgo BBHs in the first and second observing runs have been consistent with having a zero effective spin, and only a few events show non-zero effective spin $\chi_{\rm eff}$ \citep{Abbottetal:2018vb,Belczynski:2017wa,Roulet:2019js}. GW151226 and GW170729 have positive values of  $\chi_{\rm eff}=0.18^{+0.20}_{-0.12}$ and $0.36^{+0.21}_{-0.25}$, respectively. Notably, \cite{Zackay2019PhRvD} discovered a highly  spinning BBH, GW151216, from the publicly available LIGO-Virgo first observing run data. Although their estimate of the probability that GW151216 is of astrophysical origin is not very high, $0.71$, the discovery of GW151216 favors the field binary scenario \citep{Piran2020ApJ}. 

The recent discovery of the asymmetric BBH merger, GW190412 \citep{Abbottetal:2020tl}, is interesting because of two (perhaps related) facts: 
(i) It is a low FAR event with a non-zero effective spin of $\chi_{\rm eff}=0.28^{+0.07}_{-0.08}$, which provides the clue to the formation of spinning BBHs. 
(ii) It has a low mass ratio of $q=0.25^{+0.06}_{-0.04}$. This second fact is interesting on its own as dynamical formation channels would have a hard time accounting for such mergers, leaving field formation and its variants as a more promising channel to account for such mergers.

The LIGO/Virgo Scientific Collaboration (LVC) has reported that the dimensionless spin magnitude of the primary is large $\chi_1=0.46^{+0.12}_{-0.15}$, while that of the secondary is unconstrained.
\citet{Olejak:2020wj} claim to have produced such mergers through conventional binary population synthesis where $\sim $10\% of local BBH mergers in their model have $q<0.41$.
We note that in \citet{Olejak:2020wj}, the spin of the primary is largely determined at birth and matter accretion onto the BH in the later evolutionary phase of the binary only increases the spin magnitude to within a few percent.
Recent work by \citet{Mandel:2020vx} suggest that the spin of the secondary can instead be large while the primary is non-spinning and find slightly different results, $q=0.31^{+0.05}_{-0.04}$ and $\chi_{\rm eff}=0.20^{+0.03}_{-0.04}$.
In this scenario, the spin of the secondary originates from a tidal spin-up process \citep{Kushnir2016MNRAS, Zaldarriaga:2017fn,HP2017ApJ}. 

According to a model investigated by \citet{Olejak:2020wj} in which the overall BH-BH merger rate density in the local universe is: $73.5\rm Gpc^{-3} yr^{-1}$, for systems with $q<0.21$, 0.28, 0.41, and 0.59 the rate density is estimated to be: 0.01, 0.12, 6.8, and 22.2 $\rm Gpc^{-3} yr^{-1}$, respectively.
Therefore, if we believe the median mass ratio $q=0.25$ for GW190412, binary population predicts a local merger rate of less than 0.1 $\rm Gpc^{-3} yr^{-1}$.
One can note that the predicted merger rates are susceptible to the assumed mass ratio, and it changes by two orders of magnitude from systems with $q<0.21$ to systems with $q<0.41$. 
However, the situation is even worse with the predicted rates: The quoted rates from \citet{Olejak:2020wj} based on the mass ratio does not impose any other cuts on the primary mass, secondary mass, and the effective spin. 
Including all these cuts in the mass and spin parameter space reduces the predicted rates for $q<0.41$ sample with overall merger rate of 6.8 $\rm Gpc^{-3} yr^{-1}$ to 0.11 $\rm Gpc^{-3} yr^{-1}$ \citep{Olejak:2020wj}.

Separate from the mass ratio, the effective spin of this event further restricts formation scenarios:
If we believe the primary BH's spin is large, then the formation mechanisms that suggest BHs are born slowly rotating \citep{Fuller:2019gc,Fuller:2019jz} need to be re-visited \citep{baibhav2020mass,Safarzadeh:2020gs}.
If, on the other hand, we believe the secondary is providing the spin budget of the system, given that such systems have inherently short merging timescales (because of the very small separations which are necessary for tides to be effective), 
their local merger rate would be determined by the local star formation rate of the universe.

However, different groups adopt different assumptions for the BH spin at birth, and assume secondary astrophysical processes, besides tides, that lead to high effective spin BBH mergers. For example, the rates quoted in \citet{Olejak:2020wj} are based on models with mild spins for BHs at birth. Therefore, the primary BH's spin lies in the range of ($\chi_1 = 0.17-0.59$) consistent with LIGO's estimate. Consequently, even lower merger rates would be predicted if \citet{Olejak:2020wj} had implemented high efficiency angular momentum transport schemes \citep{Fuller:2019gc,Fuller:2019jz}.

But what is the merger rate of GW190412 like systems? There are two simple ways to approximately compute it: the merger rate of BBHs inferred from 
LIGO's O1 and O2 runs is in the range of 10-100 $\rm Gpc^{-3} yr^{-1}$ \citep{Abbottetal:2018vb} out of which one event is detected with such mass ratio. This leads to the observed merger rate of about  $\sim$ 1-10~ $\rm Gpc^{-3} yr^{-1}$ for such systems.
A more careful calculation could be done to estimate the merger rate of such systems: 
the LIGO-Virgo detection horizon for BNS mergers during the O3 observing run is $\approx 135$ Mpc \citep{Abbottetal:2020uq}. The detection horizon scales as $\propto M_c^{5/6}$ where $M_c$ is the chirp mass of the system (this is an approximation, and 
in detail, it would depend on the mass ratio and spin of the systems).
The chirp mass for GW190412 is $M_c \approx 15 \msun$, which makes the detection horizon for such a system to reach about 1 $\rm Gpc$, from which one can compute the volume $V$.
Assuming observing time of approximately $T\approx$50 days for O3, 
One can estimate a merger rate of GW190412 type system following Eq (17) of \citet{Abbottetal:2016gj}:

\be
  p\left(R|\mathrm{GW190412}\right) \propto \\
  \sqrt{R \langle VT \rangle}
  \exp\left[ - R \langle VT \rangle \right],
\ee

 to be $\sim 1.7^{+2.5}_{-1.4}~\rm Gpc^{-3} yr^{-1}$ (90\% confidence interval). 
Although these are rough estimates, one can, however, compare to the predicted merger rate of such a system from population synthesis models. 
Therefore, if indeed the actual value of the mass ratio is $q<0.21$, or $q<0.28$, then the predicted rates are in tension with the observed event rate. 

We note that the observing time used in this calculation, to be 50 days, is an average guess and we will not know the actual rate of this event until the O3a catalog is out. One extreme choice is $T=12$ days (since GW190412 was detected 12 days into O3a). 
This would result in a merger rate of $\sim 6.8^{+10.5}_{-5.2}~\rm Gpc^{-3} yr^{-1}$ (90\% confidence interval). 
On the other hand, one can estimate the observing time by taking all of O3a ($\approx180$ days) multiplied by some assumed duty cycle (for example, 0.5), 
which would result in a merger rate of $\sim 1.2^{+1.4}_{-1}~\rm Gpc^{-3} yr^{-1}$ (90\% confidence interval). 
This also would assume that GW190412 is the only event found in its class in O3a. 
Therefore, one can argue from the total uncertainty budget than the merger rate is confined within $0.2-17.3~\rm Gpc^{-3} yr^{-1}$ (90\% confidence interval).

In this \emph{Letter} we emphasis on this point, that merger rate of tidally locked binaries in the local universe is too low compared to the observed values of such systems, 
and therefore suggest that it may be that our assumptions regarding the spin of the BHs at birth need to be re-visited. 
The structure of this \emph{Letter} is as follows: in \S \ref{sec:td} we explain our model of spin-up process in detail. In \S \ref{sec:result} we present our predicted merger rate of tidally locked systems in the local universe, and 
in \S \ref{sec:summ} we summarize our results and conclude. 

\section{Tidal spin-up model}\label{sec:td}
It is possible that the spin of the secondary BH originates from the tidal spin-up of its progenitor before the collapse into a BH.  Such processes, however, would require the initial separation between the progenitor of the secondary BH, and the 
primary BH to be small. The consequence of this is a short delay time for such binaries. The short delay time would not be a problem by itself; however, the local merger ($z=\mathcal{O}(0.1)$) of BBHs with a tidally spun-up secondary may be in tension with the observations: BBHs with a primary mass of $\sim 30M_{\odot}$ are formed at low metallicities, and the tidal spin-up process itself is more efficient at low metallicities. Therefore, one would expect that a 
BBH consists of a $\sim 30M_{\odot}$ primary, and tidally spun-up secondary is preferably formed at low metallicities. 
In the following,  we try to answer the question of whether or not there is a sufficiently large parameter space for local BBH mergers with a tidally spun-up secondary. 

Following the description of tidal synchronization by \cite{Kushnir2016MNRAS,Zaldarriaga:2017fn,HP2017ApJ},
we consider BH-WR binaries as the last stage of binary evolution leading to  BH-BH mergers. Here BH-WR binaries are regarded as the outcome of a common envelope phase. 
Therefore, we assume that both the initial spin parameter of the WR star and the spin parameter of the primary BH is zero.  
We compute the evolution of the  spin parameter of the WR star,  $a_*$, 
for a given mass loss rate, $\dot{m}_2$, and initial values of $m_2$ and the orbital separation, $d$, by using the analytic formula developed by \cite{Kushnir2016MNRAS}:
\begin{eqnarray}
\frac{da_*}{dt} = \frac{a_{\rm syn}}{t_{\rm syn}}\left( 1-\frac{a_*}{a_{\rm syn}}\right)^{8/3} -\frac{a_*}{t_{\rm wind}},
\end{eqnarray}
where $a_{\rm syn}$ is the spin parameter when the star is synchronized at $d$, $t_{\rm syn}$ is the tidal synchronization timescale, and $t_{\rm wind}$ is the time scale of angular momentum loss  through winds. Here we assume that the mass is lost from a spherical shell at the stellar surface and the star is in rigid rotation, corresponding to $\approx 0.1m_2/\dot{m}_2$  \citep{Kushnir2016MNRAS}. Note that $m_2$ and $d$ evolve with time due to isotropic wind mass loss but the mass and spin parameter of the primary BH, $m_1$ and $a_1$, are assumed to be constant with time. 
For WR stars, \cite{Kushnir2016MNRAS} show that the tidal synchronization timescale is given by mass ratio, $q=m_2/m_1$, and coalescence time\footnote{We define the coalescence time to be the time of the core collapse of the secondary to the merger.}, $t_c$ as
\begin{eqnarray}
t_{\rm syn} \approx 10
q^{-1/8}\left(\frac{1+q}{2q} \right)^{31/24}\left(\frac{t_c}{1\,{\rm Gyr}}\right)^{17/8}\,{\rm Myr}. \label{eq:synch}
\end{eqnarray}
 The tidal lock occurs on a time scale much shorter than the stellar evolution time scales when $t_c \ll 100\,{\rm Myr}$. The spin parameter of the secondary black hole  is given by $\chi_2={\rm min}(a_*/m_2,1)$, where $a_*$ is the WR's spin angular momentum at the end of its life. Since the direction of the spin angular momentum is parallel to the orbit vector the effective spin is simply $\chi_{\rm eff}=m_2 \chi_2 /(m_1+m_2)$.

In order to relate the mass loss rate of WR stars to metallicity, we make an assumption that the mass loss rate can be described by the following form
\begin{eqnarray}
\dot{m_2}(m_2,Z) = f_{w}\dot{M}_{0}\left(\frac{m_2}{M_{\odot}} \right)^{\alpha}\left(\frac{Z}{Z_{\odot}} \right)^{\beta},\label{eq:loss}
\end{eqnarray}
where $Z$ is the metallicity at ZAMS. Here we fix $\dot{M}_0 = 10^{-5.73}M_{\odot}/{\rm yr}$ and $\alpha\approx 0.88$  based on the result for the sample of Galactic WN plus WC stars obtained by  \cite{Nugis2000A&A}.
For our fiducial model, we choose $\beta=0.8$ and $f_{w}=1$. In reality, the index $\beta$ evolves with time because the metallicity at the surface of a WR star increases with time  (see,  e.g., \citealt{Eldridge2006A&A,Yoon2017MNRAS} for detailed studies). However,  it turns out that our results only weakly depend on $\beta$ for the range of $0.5\leq \beta\leq 1.1$, and hence,
we use $\beta=0.8$ in the following.

To obtain the distribution of the spin parameters of BBH mergers at a given redshift, we integrate the BBH merger rates arising from different metallicities and redshifts. 
We assume the formation rate of merging BBH is proportional to the cosmic star formation rate \citep{Madau:2014gt}:
\begin{eqnarray}
\dot{\rho}(z)=\frac{\dot{\rho}_0(1+z)^{2.7}}{1+((1+z)/2.9)^{5.6}},\label{eq:his} 
\end{eqnarray}
where $z$ is redshift and $\dot{\rho}_0$ is the present-day star formation rate. We also assume that the fraction  of star formation at and below metallicity mass fraction of $Z$ at given redshift is described by \citep{Langer2006ApJ}:
\begin{eqnarray}
\Psi(z,Z) = \frac{\hat{\Gamma}(0.84,(Z/Z_{\odot})^210^{0.3z})}{\Gamma(0.84)},\label{eq:z} 
\end{eqnarray}
where $\hat{\Gamma}$ and $\Gamma$ are incomplete and complete gamma-functions, respectively. At higher metallicities, the mass of the primary at the ZAMS is larger, and therefore the formation efficiency of more massive stars is reduced according to the initial mass function. Here we use a relation between the initial mass of a zero-age WR star to the mass at the ZAMS \citep{Limongi2018ApJS}. Then we assume that the distribution of the primary masses at the ZAMS follows the Kroupa 2001 initial mass function \citep{Kroupa:2001ki}. 

\begin{figure}
\centering
\includegraphics[width=0.45\textwidth]{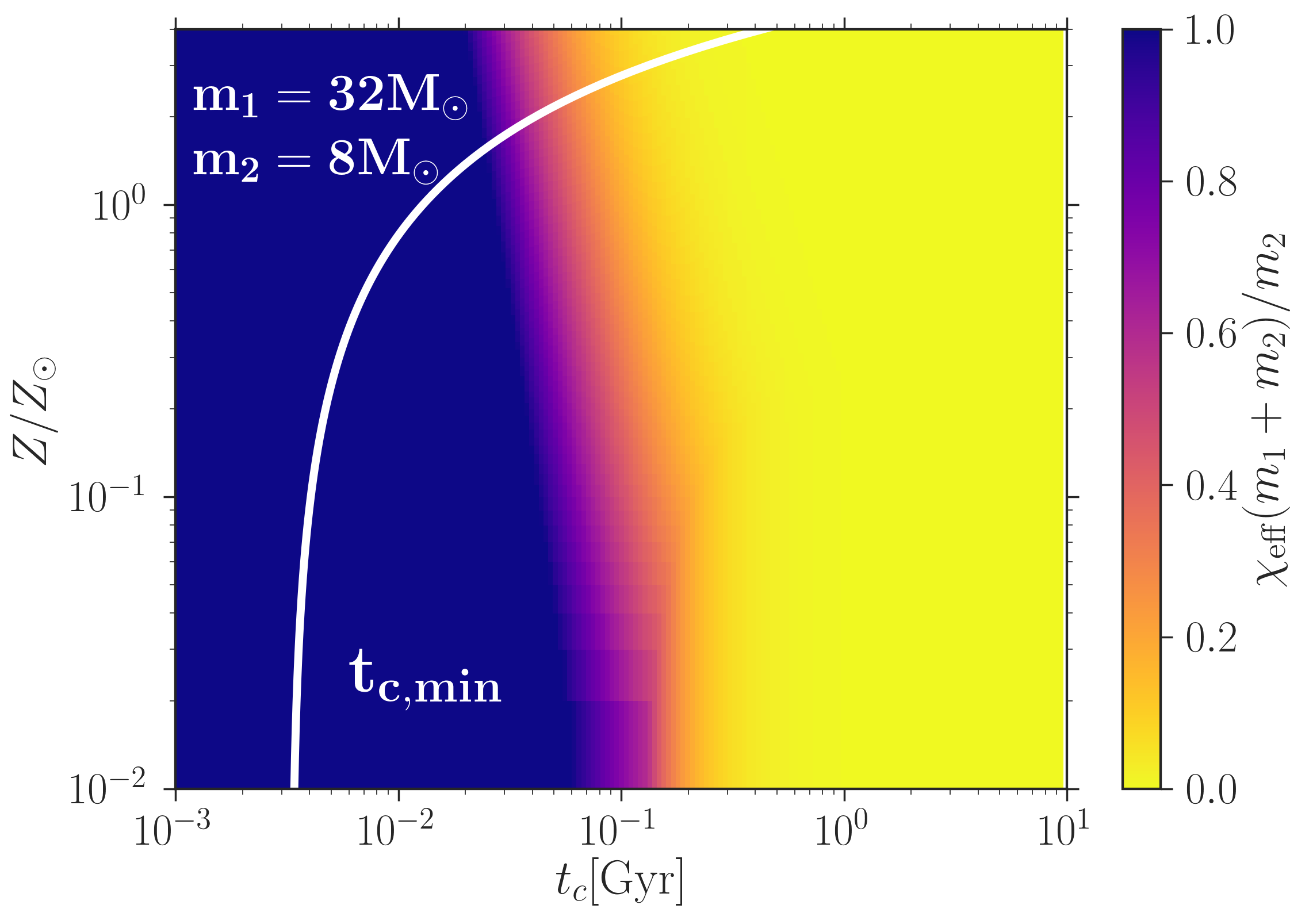}\\
\includegraphics[width=0.45\textwidth]{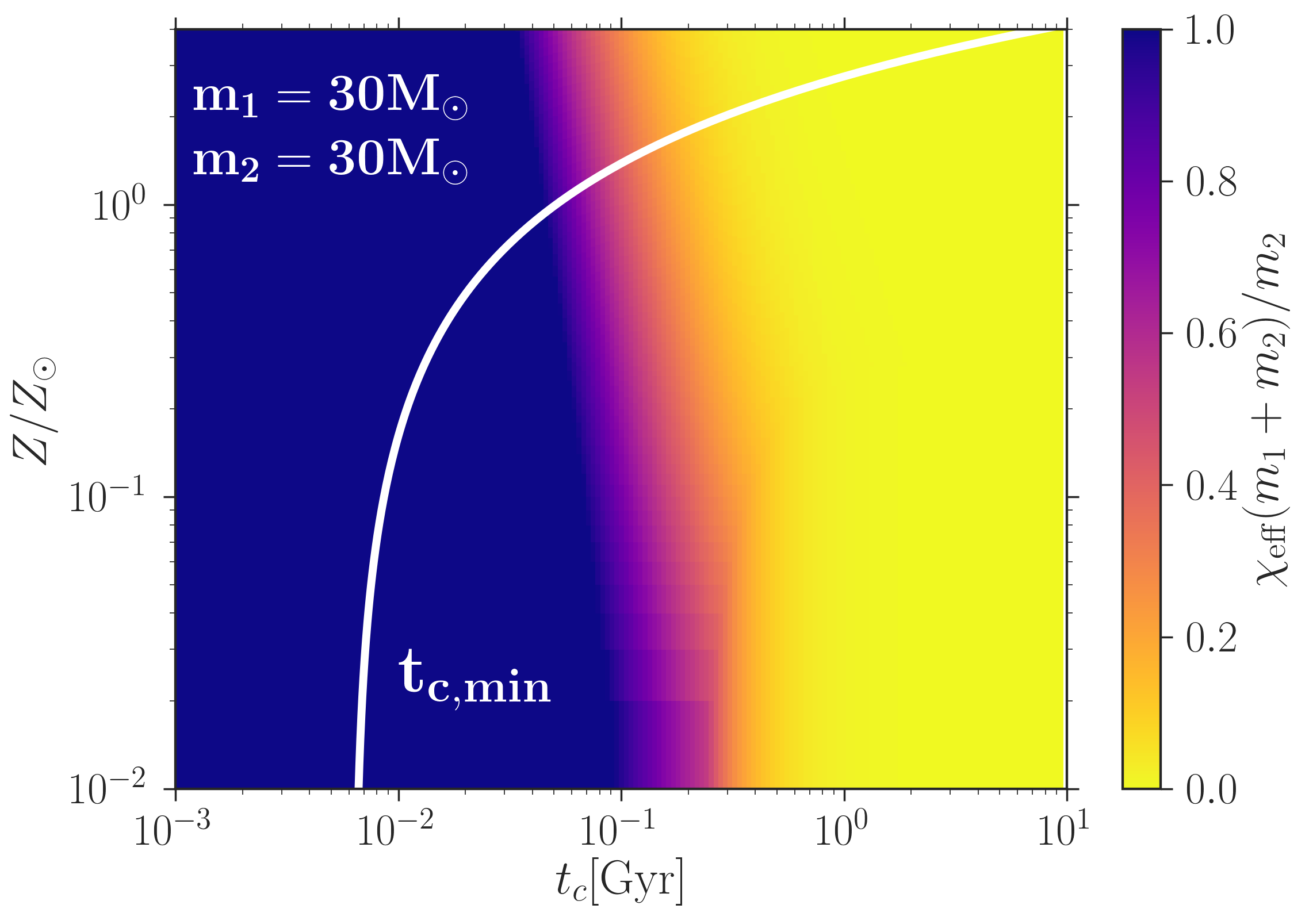}
\caption{Spin parameter of the secondary black hole as a function of metallicity and coalescence time. {\it Upper} and {\it lower} panels show $32$-$8M_{\odot}$ and $30$-$30M_{\odot}$ cases, respectively. A curve labeled by $t_{c,{\rm min}}$ is the minimum coalescence time estimated by assuming that the Roche radius at the beginning of the BH-WR binary phase equals to the stellar radius. We note that in this calculation we have assumed non-spinning primary BHs.
 }\label{fig:tc_Z}
\end{figure}

Finally, we use the delay time distribution of BBH mergers 
\begin{eqnarray}
DTD(t_c)= \begin{cases} C_0 t_c^{-\gamma}&\mbox{for }t_c>t_{c,{\rm min}}, \\ 
0 & \mbox{otherwise}  \end{cases} \label{eq:t} 
\end{eqnarray}
where $C_0$ is a normalization constant  and we use $\gamma=1$ for the fiducial model, which is motivated by the semi-major axis distribution of massive binaries (e.g., \citealt{Sana:2012gu})\footnote{\cite{Moe:2017eg} found correlations between orbital periods, masses, and eccentricities of stellar binaries. However, it is still unclear how these correlations affect the delay time distribution of BBHs through binary evolution \citep{Klencki2018A&A,Stanway2020MNRAS}. }. Later we also study a case with $\gamma=1.5$. 
Note that, however,  such a steep delay time distribution requires an extremely steep distribution in the  orbital separation at BBH formation.

\section{Results}\label{sec:result}
\begin{figure}
\centering
\includegraphics[width=0.45\textwidth]{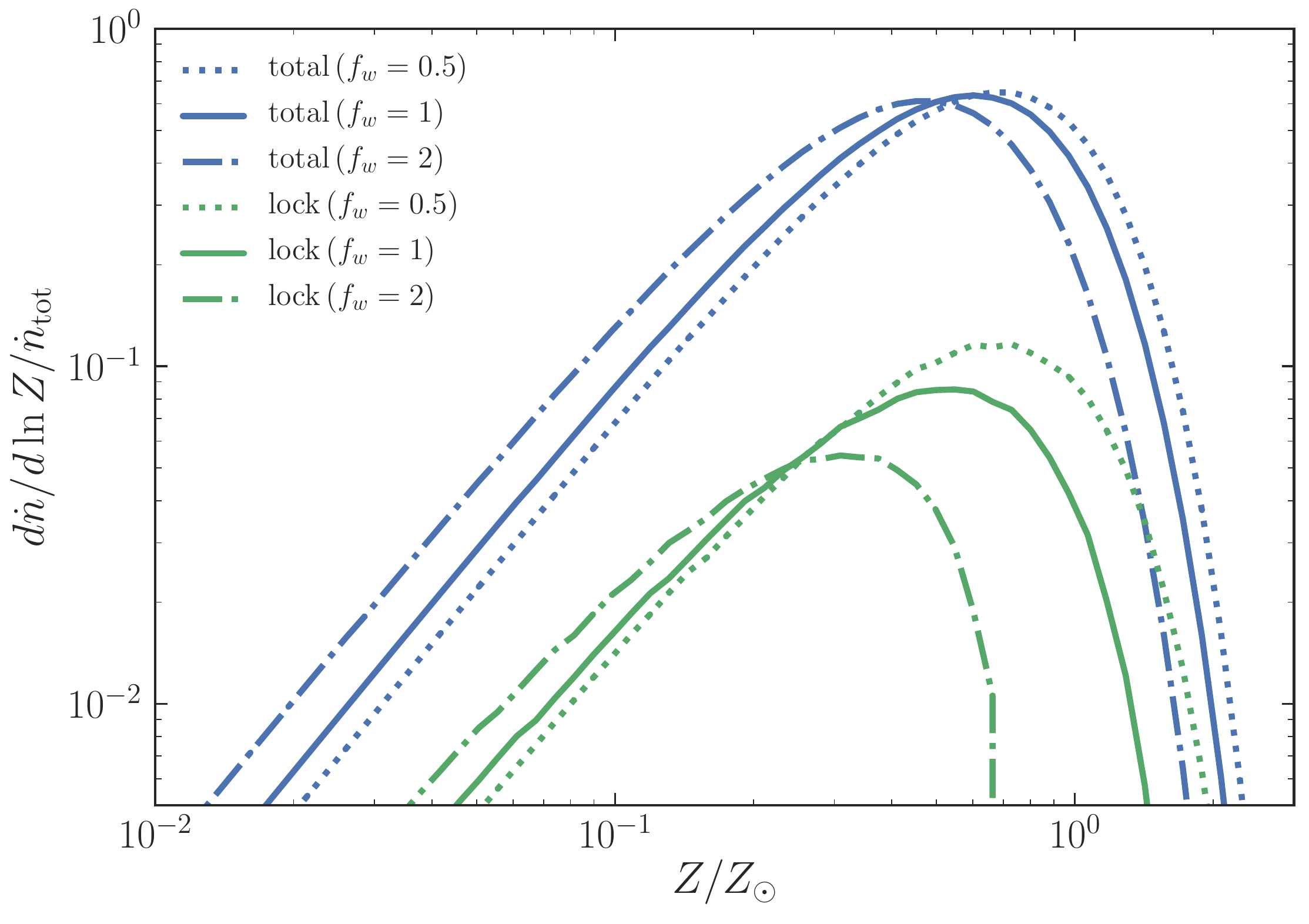}\\
\includegraphics[width=0.45\textwidth]{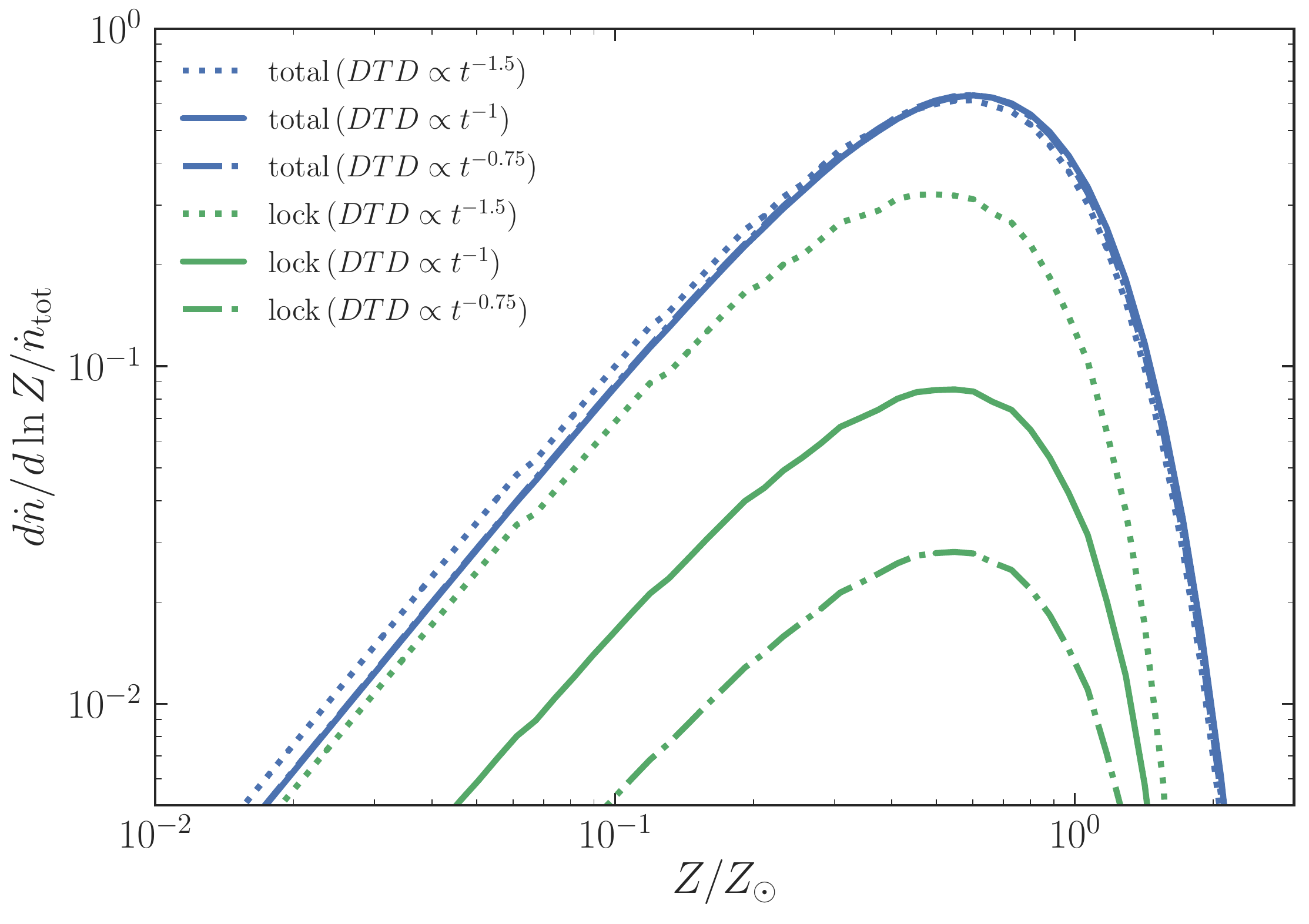}
\caption{Normalized distribution of BBH mergers
with $32$-$8M_{\odot}$ at $z=0.15$ as a function of metallicity. We denote by ``locked" BBH mergers with $\chi_{\rm eff}(m_1+m_2)/m_2>0.8$. {\it Upper} and {\it lower} panels show the dependence on the strength of wind mass loss, $f_w$, and the exponent of the delay time distribution, $\gamma$, respectively. For our fiducial model, $f_w=1$ and $\gamma=1$, the fraction of locked mergers is $\sim 15\%$ after integrating the rates over the metellacity. This fraction decreases roughly as $\propto f_w^{-1}$.
Increasing the mass loss rate reduces the locked fraction particularly at higher metallicities. The integrated fraction of locked mergers is sensitive to the $\gamma$ value;
$5$, $15$, and $50\%$ for $\gamma=0.75$, $1$, and $1.5$, respectively.}\label{fig:merger}
\end{figure}

Figure \ref{fig:tc_Z} shows the spin parameter of the secondary black hole, $a_2/m_2$, as a function of coalescence time $t_c$ and the metallicity $Z$.   The distribution of $a_2/m_2$ values depends very weakly on the mass ratio. 
The coalescence time required for BBHs composed of a tidally spun-up 
secondary is $\lesssim 100\,{\rm Myr}$, which is much shorter than the typical lookback time of BBH mergers.
Also depicted by a curve in figure \ref{fig:tc_Z}  is the minimum coalescence time, $t_{c,{\rm min}}$, derived based on the assumption that the  semi-major axis at the beginning of the BH-WR binary phase must be larger than the one at which the stellar radius equals the Roche radius. This may be a good estimate of  $t_{c,{\rm min}}$ unless the secondary ejects a fraction of its mass or receives a significant kick at the core collapse. With this assumption, the $t_{\rm c,min}$ value becomes larger at higher metallicities because the mass loss effect on the semi-major axis is more significant at higher metallicities, but in general, there is a reasonably large parameter space for a tidally spin-up secondary. Note that the $t_{\rm c,min}$ values are 
smaller for the $32$-$8M_{\odot}$ case than the equal mass case. The reasons are that the  semi-major axis at the Roche radius decreases with mass  and that the mass loss effect on the orbit is less significant for asymmetric binaries. Consequently, 
the parameter space where a BBH consists of a tidally-locked secondary is slightly larger for more asymmetric BBHs. 

The fraction of BBH mergers consisting of a tidally-locked secondary is obtained by comparing the area of the high spin region of figure \ref{fig:tc_Z} to the total area with a weight function that takes into account: (i) the cosmic BBH formation history (equation \ref{eq:his}), (ii) cosmic metallicity evolution (equation \ref{eq:z}), and (iii) delay time distribution of BBH mergers (equation \ref{eq:t}). 
Figure \ref{fig:merger} shows the normalized  BBH merger rate at $z=0.15$   as a function of the metallicity at their formation in the case of $32$-$8M_{\odot}$. We find that the merger rates sharply drop around the solar metallicity, which is consistent with the results of binary population synthesis \citep{Eldridge2016MNRAS}. The fraction of  BBH mergers with a highly spinning secondary, $\chi_{\rm eff}(m_1+m_2)/m_2>0.8$, strongly depends  on the exponent of the delay time distribution, $\gamma$.

\begin{figure}
\centering
\includegraphics[width=0.45\textwidth]{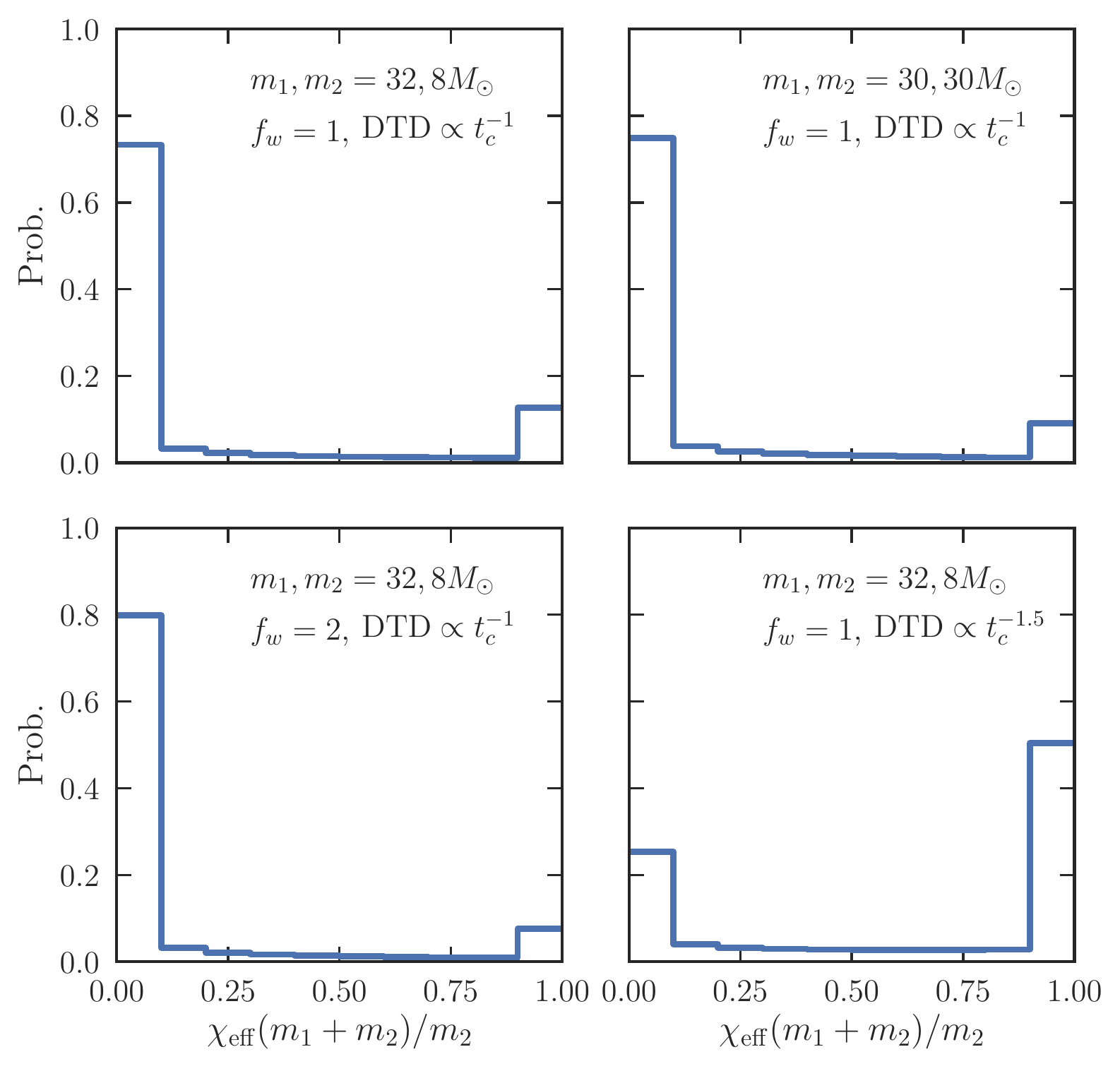}
\caption{Probability distribution of the secondary spin parameter $a_2/m_2$ for BBH merger at $z=0.15$. 
The bimodal distribution in all panels is indicative of BBHs that either undergo a tidal spin-up phase or, due to their large initial binary separation, do not experience a tidal spin-up phase. Those BBHs experiencing a tidal spin-up  phase emerge as high effective spin BBH mergers, and those that do not experience a tidal spin-up phase merge with near-zero effective spin parameter. \emph{Top left}: shows the case for a $m_1,m_2=32,~8~\msun$ system with fiducial values for wind mass loss and delay time distribution. A nearly 10-15\% of the BBHs experience a tidal spin-up process in our fiducial model. \emph{Top right}: The same is shown for an equal mass ratio binary. A nearly similar result is obtained, indicating that the mass ratio plays a minor role in our results to first order. \emph{Bottom left}: Increasing the mass loss rate by a factor of 2. This decreases the tidally spun-up binaries in that more angular momentum is carried away through the winds, and the final spin is subsequently dropped. However, the change in the ratio of the tidally spun-up systems over the total population remains similar to the fiducial case. \emph{Bottom right}: Changing the underlying delay time distribution to a steeper functional form. This will increase the fraction of the tidally spun-up systems to about 50\% and remains a parameter that our model is most sensitive to.}\label{fig:prob}
\end{figure}

We present the probability distribution of $\chi_{\rm eff}(m_1+m_2)/m_2$ values for BBHs merging at $z=0.15$ in figure \ref{fig:prob}.  With the fiducial parameters, we find 
$15\%$ and $10\%$ of mergers have a tidally spun-up secondary for $30$-$8M_{\odot}$ and $30$-$30M_{\odot}$ cases, respectively. 
The dependence of the fraction of tidally spun-up BBH mergers on the mass ratio is rather weak. We also find that total mass dependence is also quite weak. Increasing the wind mass-loss rate $f_w$ results in the smaller fraction  roughly as $\propto f_w^{-1}$.  Thus, we conclude that $10$--$15\,\%$ of BBH mergers with $q\approx 0.25$ at $z\approx 0.15$ have a tidally spun-up secondary within  reasonable ranges of the model parameters. 
Finally, we find that the high spin fraction is significantly high if the delay time distribution is very steep, e.g., $50\%$ when ${\rm DTD}(t_c)\propto t^{-1.5}_c$.

Based on the tidal spin-up model presented in this work, we can calculate the expected rate of GW190412 type systems:
\be
\mathcal{R}_{GW190412}=\mathcal{R}_{\rm BBH} f_{\rm tidal} f_{m_1,m_2},
\ee
where $\mathcal{R}_{\rm BBH}$ is the local BBH merger rate estimated to be in the range of 10-100 $\rm Gpc^{-3} yr^{-1}$. 
$f_{\rm tidal}$ is the tidal fraction of systems with mass ratio and masses similar to GW190412, which we compute to be about 10\%.
$f_{m_1,m_2}$ is the fraction of the BBH mergers with primary mass $m_1=30\msun$, and secondary mass $m_2=8\msun$. 
To compute this we assume a primary BH mass function following $p(m_1)\propto m_1^{-\alpha}$ bounded between 5 $\msun$ and 
50 $\msun$ (to account for both the lower mass gap limit of the BHs \citep{Farr:2011ct}, and upper limit due to pair-instability supernovae \citep{Woosley:2017dj}). We set $\alpha=2$ and compute the fraction of BHs with mass above 30 $\msun$ over the total population resulting in about 15\% (assuming $\alpha=1$ would slightly increase the fraction to 20\%). Furthermore, assuming flat distribution for the mass function of the secondary, the fraction of systems with $q<0.25$ equals 0.25. The combination of 
these two makes $f_{m_1,m_2}\approx0.04$. 
Therefore, the expected rate based on first-order calculation suggests  $\mathcal{R}_{GW190412}\approx0.03-0.3~\rm Gpc^{-3} yr^{-1}$. 
The upper limit on this rate agrees with the lower limit of the observed event merger rate, which is indicative of a strong tension.

With a steeper delay time distribution $t_c^{-1.5}$, 
the expected rate of GW190412-like events can be  consistent with the observed rate. Interestingly, the distribution of galactic binary pulsars indicates  such a steep delay time distribution for binary neutron star mergers \cite{Beniamini2019MNRAS}.  However, BBH mergers with $q\sim 1$ cannot have a steep delay time distribution because it predicts too many highly spinning equal-mass BBH mergers. Thus, a high rate of GW190412 may suggest that the steepness of the delay time distribution depends on $q$. 
Note that such models would be detected through stochastic gravitational background as they would predict a very high background level \citep{safarzadeh2020constraining}.

\section{summary and discussion}\label{sec:summ}

GW190412 is unusual due to its two potentially related observed facts: i) low mass ratio ($q\approx 0.25$), ii) relatively high effective spin of $\chi_{\rm eff}=0.28^{+0.07}_{-0.08}$. 
We study a field binary origin scenario in which the tidally spun-up secondary dominates the effective spin. This scenario works effectively at low metallicities and predicts a coalescence time of $\lesssim 100$ Myr. 
We explore  $\chi_{\rm eff}$ values as a function of metallicity and delay time by using an analytic model developed by \cite{Kushnir2016MNRAS}. In this model, we calculate the spin distribution at the local universe ($z=0.15$), taking into the cosmic star formation history and metallicity evolution with redshift. 
Under the assumption that the minimum delay time is set by the orbital separation at which the radius of a Wolf-Rayet secondary equals the Roche radius, we find that:
\begin{itemize}
    \item A bimodal distribution of $\chi_{\rm eff}$ with peaks at $\chi_{\rm eff}\sim 0$ and $\chi_{\rm eff}(m_1+m_2)/m_2\sim 1$ is a robust consequence of tidal spin-up models. This  is consistent with the previous results of \cite{Zaldarriaga:2017fn,HP2017ApJ,Piran2020ApJ}.
    \item $10$--$15\%$ of BBH mergers with $q\approx 0.25$ have a nearly maximum spinning secondary. Thus, we expect more asymmetric mergers with $\chi_{\rm eff}\sim 0$ to be detected in the future observing runs. This fraction for BBH mergers with $q\approx 0.25$ is by a factor of $\sim 1.5$ larger than the equal mass case. 
    \item A steep delay time distribution such as $DTD\propto t_c^{-1.5}$
    results in the mergers composed of a tidally spun-up secondary that would dominate over non-spun-up mergers.
\end{itemize}

Another important observational feature of GW190412 is the marginal detection of precession, suggestive of a non-zero in-plane spin \citep{Abbottetal:2020tl}. To obtain a non-zero in-plane spin
that is sufficiently large to induce observable precession, for any field binary scenario,  requires a black hole natal kick of which magnitude is comparable to the orbital velocity at the collapse.  For the tidal spin-up model, this corresponds to an extreme natal kick, $\gtrsim 500\,{\rm km/s}$ in the case of $\chi_{\rm eff}(m_1+m_2)/m_2\sim 1$.
Such large natal kicks are not expected from observations of Galactic X-ray binaries. For instance, the natal kicks for Cyg X-1 and GRO J1655-40 are constrained to $\lesssim 80\,{\rm km/s}$ and $\lesssim 120\,{\rm km/s}$, respectively \citep{Wong2012ApJ,Willems2005ApJ}.  
Therefore, if non-zero $\chi_p$ of GW190412 is really significant, it disfavors the tidal spin-up model. However, such a strong natal kick perpendicular to the orbital plane may result from the launch of a one-side jet from a BH-accretion disk system \citep{Barkov2010MNRAS}.  We note that the formation of the accretion disk and mass ejection naturally occur at the collapse of tidally spun-up secondary because the spin parameter of the collapsing star may exceed unity \citep{batta2019accretion}.

While LVC's analysis results in the primary BH to be spinning with no constraint on the secondary's spin, if we assume GW190412 is a tidal spin-up system, the spin of the binary will be largely attributed to the secondary BH.  
The tidal spin-up scenario faces two challenges: 
1) The predicted merger rate of such system may be too low in this formation channel. \citet{Olejak:2020wj} predicts less than 0.1 $\rm Gpc^{-3} yr^{-1}$ for binaries with $q<0.28$, 0.11 $\rm Gpc^{-3} yr^{-1}$ for binaries
with $q<0.41$ and similar mass and spin to GW190412. This is similar to our results presented in this work. However, the observed merger rate of such systems is about ten times more common than the predictions suggesting a large tension between the tidal spin-up model for this event.
2) This formation channel can not explain the possible in-plane spin of GW190412 if we assume the spin direction of the BHs at birth are aligned with their orbital angular momentum vector. 
Imposing random natal kicks on the BHs at birth to account for the observed precession of GW190412 would only exacerbate the tension in the estimated local merger rate and the observed high merger rate for such systems. 
This caveat applies to both our model and those presented in \citet{Olejak:2020wj}.

Any successful model attempting to explain GW190412 should take into account the rather high ($1.7^{+2.5}_{-1.4}~\rm Gpc^{-3} yr^{-1}$) local merger rate of this system. 
Separate from the tidal spin-up channel that we challenge in this work, dynamical formation channels of the BBHs \citep{Samsing_2018,Rodriguez:2019fx} would have a difficulty predicting such high merger rate for such low mass ratio systems. 
For example, \citet{gerosa2020astrophysical} argue that second-generation mergers in globular clusters are exceedingly rare, and one might need super dense environments such as 
super star clusters or AGN disks \citep{Bartos_2017} to account for such events. 
Other scenarios based on, for example, the quadruple configuration for the birth of such systems \citep{hamers2020gw190412}, although successful in reproducing all the observables of GW190412, would imply that birth rate of the quadruples are high in the universe which would encourage a more scrutinized look into the birth rate of such systems. 

Lastly, we note that the merger rate estimates based on the local star formation rate at a specific metallicity bin are largely uncertain \citep{Chruslinska:2018bv} and model-dependent \citep{Lamberts_2016}. 
Therefore, conclusions regarding the formation channel of GW190412 is tied to our prior assumptions about the local star formation rate in the universe. Including all sources of uncertainty is beyond the scope of this \emph{Letter} and 
we defer such detailed analysis to future works.

\acknowledgements We are thankful to the referee for their detailed comments. We are also thankful to Michael Zevin, Avi Loeb, Tsvi Piran, and Emanuele Berti for comments on the earlier version of this manuscript.
This work is supported by the National Science Foundation under Grant No. AST-1440254.
\bibliographystyle{yahapj}
\bibliography{the_entire_lib.bib}
\end{document}